\documentclass[aps,prl,preprint]{revtex4}
\usepackage{graphicx}

\begin{document}

\noindent \textbf{In response to ``the comments by Murphy et al.''}
\par\noindent {\it --R. Srianand$^1$, H. Chand$^2$, P. Petitjean$^2$ \& B. Aracil$^2$}
\par\noindent {\it${^1}$ IUCAA, Postbag 4, Ganeshkhind, Pune 411001, India} 

\par\noindent{\it $^2$ 
Institut d'Astrophysique de Paris, 98bis Boulevard Arago, 7501, Paris, France}
\par\noindent
In their comment, Murphy et al. criticize the fitting procedure
we used in two previous papers [Srianand et al. 2004 (Paper I) and 
Chand et al. 2004 (Paper II)]  and conclude that Paper I offers no 
stringent test to previous evidence for varying fine structure constant. 
We think this is a hasty conclusion as 
(a) our procedure is robust as shown in Paper II; (b) the data used 
by Murphy et al., in particular the error array, are different from ours and
there are 
differences in the fitting procedure; 
(c) despite these differences, 70\% of their individual measurements are 
consistent with that quoted in Paper II.

\begin{figure}
\begin{center}
\includegraphics[bb= 18 418 584 716,width=8.cm,height=3.5cm,clip=true]{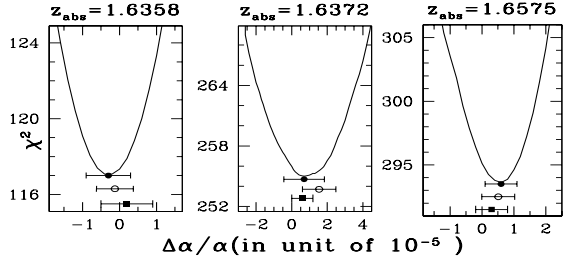}
\end{center}
\vspace{-2em}
\caption{
Examples from our reanalysis (see Point 1).
}
\label{fig2}
\end{figure}

\noindent{\bf Point 1:} In Paper II, we explain in details
our procedure. In particular, we used $\Delta\alpha$/$\alpha$ as an external 
parameter (as in Webb et al. 1999) when Murphy et al. used it as 
an additional fitting parameter.  This choice has been tested extensively using simulations. 
To reiterate this point, we have refitted the systems in our sample 
using VPFIT keeping $\Delta\alpha/\alpha$ as an external parameter 
(three examples are given in Fig.~\ref{fig2}). Our new results 
(filled circles) match our  original results (squares) and Murphy et al.'s (open squares) 
within $1\sigma$.  We point out that fluctuations in $\chi^2$ curves get indeed 
smoothed after a large number of iterations but the results from the first and last 
iterations are found to be very similar.  We find significant differences  in only 
two cases (see below) compared to our earlier results.

\begin{figure}
\begin{center}
\includegraphics[bb= 18 433 319 698, width=7.cm,height=5.cm,clip=true]{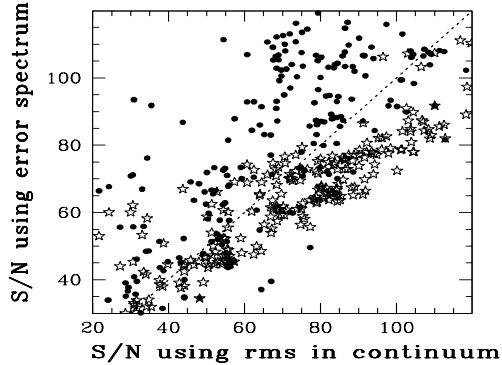} 
\end{center}
\vspace{-2em}
\caption{
SNR (1/$\sigma$) used in Murphy et al. 
(dots) and our analysis (stars) given versus SNR 
in the continuum (Point 2).
}
\label{fig1}
\end{figure}

\noindent{\bf Point 2:}  
In the course of the analysis presented in Paper II, we realized that the 
treatment of errors is not trivial. Therefore we defined two errors for each pixel: 
one is the error calculated by the ESO-pipeline 
and one is calculated from the scatter of the different exposures (usually
more than eight exposures). 
In principle the two errors should  be of the same order but usually they are not
because of the non-trivial observational procedure.
This can be estimated by comparing the above errors with the scatter in the continuum.
Murphy et al. used the data made available on the web with standard errors.
This is illustrated in Fig.~\ref{fig1} where we show the SNR (1/$\sigma$) used by Murphy
et al. (dots) and by Srianand et al. (stars) versus the SNR as measured in the 
continuum around the absorption lines used in our analysis. 
It is apparent that errors used by Murphy et al. are underestimated. 
Ours are consistent in the low SNR regime and slightly over-estimated  
at high SNR.
It must be remembered however that these measurements are done in the continuum
and differences are more crucial in the lines (where we cannot perform this experiment).
In addition, 
our procedure takes into account the differences in  spectral resolution 
in different settings (because of different observational settings and seeing conditions), 
while this is not the case with VPFIT.
\par\noindent
\noindent{\bf Point 3:}
Despite these differences, it is clear from Fig. 3 (left panel) and 
Table~1 of Muphy et al. (2006)
that their measurements match ours (Paper II) at 
$\le 1\sigma$ level for 16 systems. The corresponding weighted mean is
$\Delta\alpha/\alpha = (+0.06\pm0.18)\times10^{-5}$. 
For the same systems Chand et al. find 
$\Delta\alpha/\alpha = (+0.03\pm0.09)\times10^{-5}$.
It is also easy to recognize that only two $>4\sigma$ deviant systems (the $z= 1.5419$ system towards 
Q~0002$-$422  [$\Delta\alpha/\alpha = (-4.655 \pm 0.988) \times 10^{-5}$] and 
the $z = 0.8593$ system towards Q~0122$-$380 [$\Delta\alpha/\alpha = (-4.803 \pm 0.941) \times 10^{-5}$]), 
dominate the final result by Murphy et al. By removing these two discrepant systems 
we get  $\Delta\alpha/\alpha = (-0.19 \pm 0.16) \times 10^{-5}$ for 21 points.  
Our reanalysis using VPFIT (see point 1) keeping constant resolution across the spectrum 
and with identical initial guess parameters leads to $\Delta\alpha/\alpha = 
(0.01\pm0.15)\times10^{-5}$ for 21 systems (excluding two systems that deviate at
more than 3$\sigma$ level)
with very little scatter ($\chi^2_\nu \sim 1$) contrary to the claims by Murphy et al. 
Thus we believe, the results presented in Paper I and II are robust 
(although errors are probably larger) and are not due to 
the failure in our fitting procedure. 

In any case, our result disagrees with earlier claims for a variation of $\alpha$
by more than 3$\sigma$.
We are now awaiting the full independent analysis of UVES data by Murphy et al. and
their results.

\end{document}